\documentclass[cits]{PoS}
\usepackage{amsmath}
\newcommand{\gsim}{\gtrsim}
\newcommand{\lsim}{\lesssim}
\title{
Charged Higgs pair production and neutrino effects on the triple Higgs
coupling}
\ShortTitle{BSM Higgs pair production}

\author{\speaker{Julien Baglio}\\
  Institut f\"{u}r Theoretische Physik, Eberhard Karls Universit\"at
  T\"ubingen,\\
  Auf der Morgenstelle 14, 72076 T\"ubingen (Germany)\\
  E-mail: \email{julien.baglio@tuebingen.de}}

\abstract{Higgs pair production is one of the primary goals of the LHC
program. Investigating the effects beyond the Standard Model (BSM) is
then of high interest. Two cases are presented to exemplify the impact
of BSM physics on Higgs pair production and on the triple Higgs
coupling: first a review on charged Higgs pair production mostly in
the context of Two-Higgs-Doublet of type II and in particular the
Minimal Supersymmetric SM, second a study of the one-loop effects of a
heavy neutrino on the triple Higgs coupling.}

\FullConference{Prospects for Charged Higgs Discovery at Colliders\\
         3-6 October 2016\\
         Uppsala, Sweden} 

\begin{document}

\section{Introduction}

Since the ATLAS and CMS collaborations at the CERN Large Hadron
Collider (LHC) discovered in 2012 a particle which properties are
compatible with the Standard Model (SM) Higgs boson
hypothesis~\cite{Aad:2012tfa,Chatrchyan:2012xdj}, the detailed study
of its properties have started. Run I results at 7-8 TeV do not
display any deviations with respect to the
SM~\cite{Aad:2015mxa,Aad:2015gba,Chatrchyan:2013mxa,Khachatryan:2014kca},
as well as the first results of the Run II at 13
TeV~\cite{ATLAS:2016hru,CMS:2016ixj,CMS:2016ilx}. The measure of the
triple Higgs coupling would allow for a direct probe of the scalar
potential that is directly responsible for the electroweak symmetry
breaking
(EWSB)~\cite{Higgs:1964ia,Englert:1964et,Higgs:1964pj,Guralnik:1964eu,Higgs:1966ev}. This
is one of the major goals of the LHC and of the future planned
colliders such as the electron-positron International Linear Collider
(ILC) or the Future Circular Collider in hadron mode (FCC-hh), a
potential 100 TeV follow-up of the LHC.

In the past few years there has been numerous studies on the
production of Higgs boson pairs, that is the main production mechanism
that allows for the measure of the triple Higgs coupling, see reviews
in Refs.~\cite{Arkani-Hamed:2015vfh,Baglio:2015wcg,deFlorian:2016spz}
for the latest theoretical developments in the SM. Beyond-the-SM (BSM)
effects on Higgs pair production have been studied in many different
frameworks and in particular in the Two-Higgs-Doublet-Model (2HDM)
(see e.g. Refs.~\cite{Baglio:2014nea,Hespel:2014sla,Haber:2015pua} and
the results collected in Ref.~\cite{deFlorian:2016spz}) and
in the minimal supersymmetric extension of the SM (MSSM) which is a
particular type of 2HDM in the Higgs sector (see for example
Ref.~\cite{Cao:2013si}). In 2HDMs there are five Higgs bosons in the
spectrum, and in particular there are two charged Higgs bosons
$H^\pm_{}$, for a review see Ref.~\cite{Branco:2011iw}. To allow for
the measure of the triple Higgs coupling involving charged Higgs
bosons it is necessary to produce pairs of charged Higgs bosons.

In the following we will give a brief overview of the theoretical
predictions for charged Higgs boson pair production at the LHC. As an
other example of BSM effects on the triple Higgs coupling we will then
present in a second part a recent study~\cite{Baglio:2016ijw} on the
impact of an heavy neutrino on the one-loop corrected triple Higgs
coupling in a simplified model. We stress that the results in the
latter case are quite generic and could be applicable in any model
containing heavy fermions that couple via the neutrino portal.

\section{Charged Higgs boson pair production at the LHC}

The Higgs sector of a 2HDM contains two Higgs doublets leading to 5
Higgs bosons in the spectrum, two of which are the charged Higgs
bosons $H^\pm_{}$. In a type I 2HDM only one of the Higgs doublets
couples to all fermions while in a type II 2HDM one doublet couples to
up-type quarks and the second doublet couples to down-type quarks and
charged leptons. In the MSSM, the Higgs sector is a type II 2HDM and
can be parameterised by two parameters, the ratio of the two vacuum
expectation values $\tan\beta = v_2^{}/v_1^{}$
and the mass of the pseudo-scalar Higgs boson $M_A^{}$. More
parameters are required in a general 2HDM. For a recent review on the
prospects for charged Higgs bosons searches at the LHC see
Ref.~\cite{Akeroyd:2016ymd}.

\begin{figure}
\centering
\begin{minipage}[c]{5cm}
\includegraphics[scale=0.4]{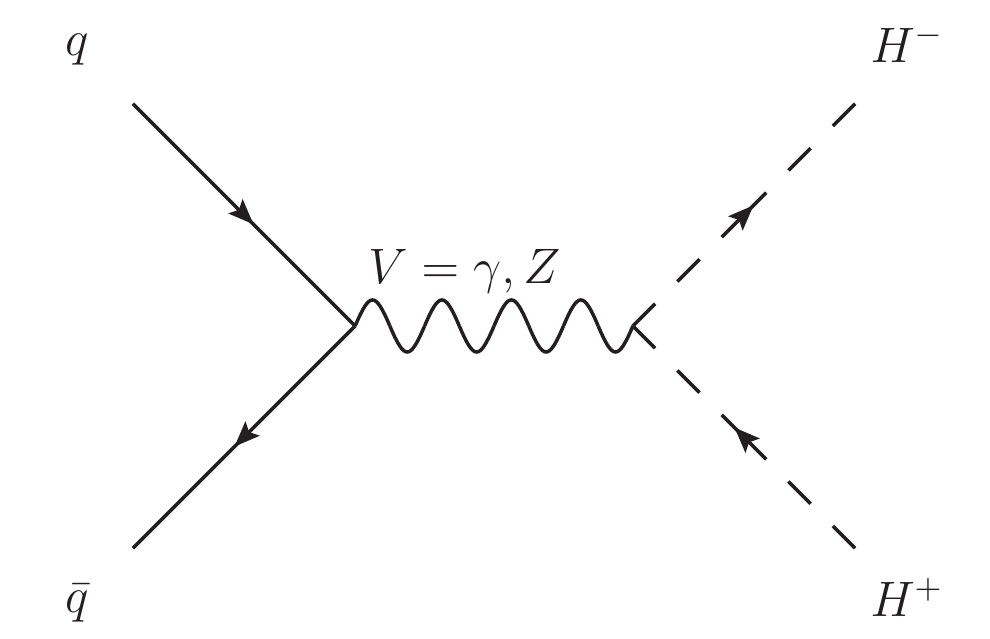}
\end{minipage}
\begin{minipage}[c]{8cm}
\includegraphics[scale=0.45]{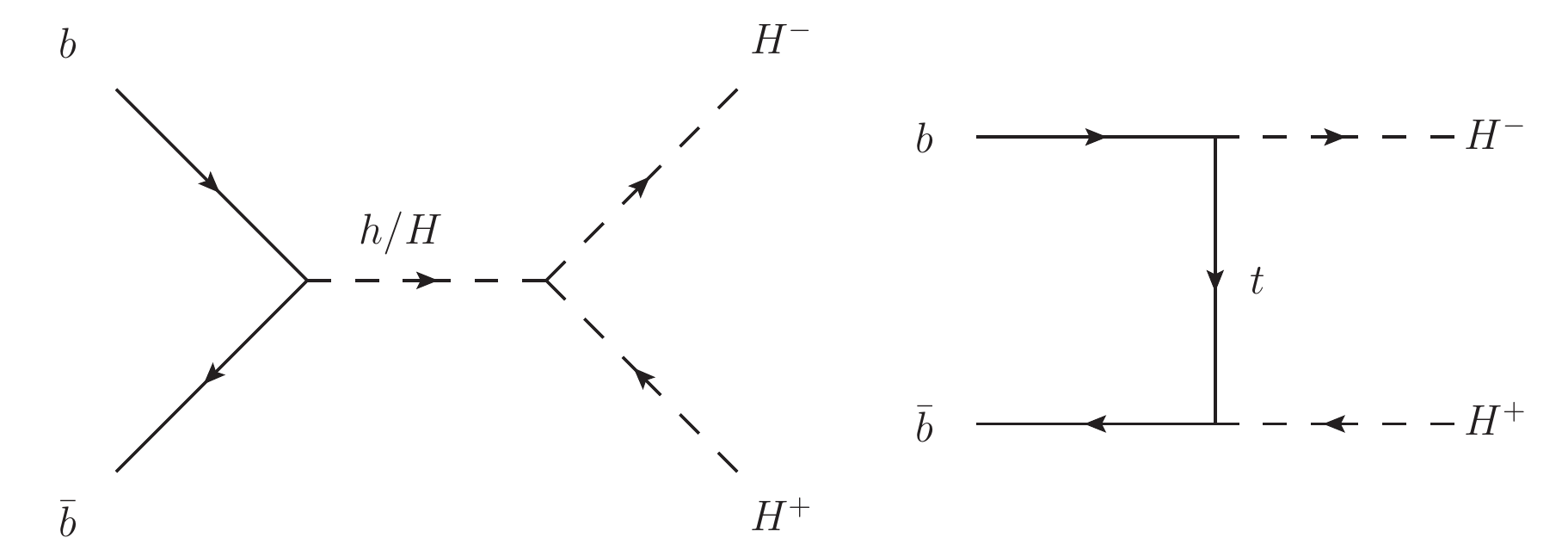}
\end{minipage}
\vspace{2mm}

\begin{minipage}[c]{10cm}
\hspace{-5mm}\includegraphics[scale=0.6]{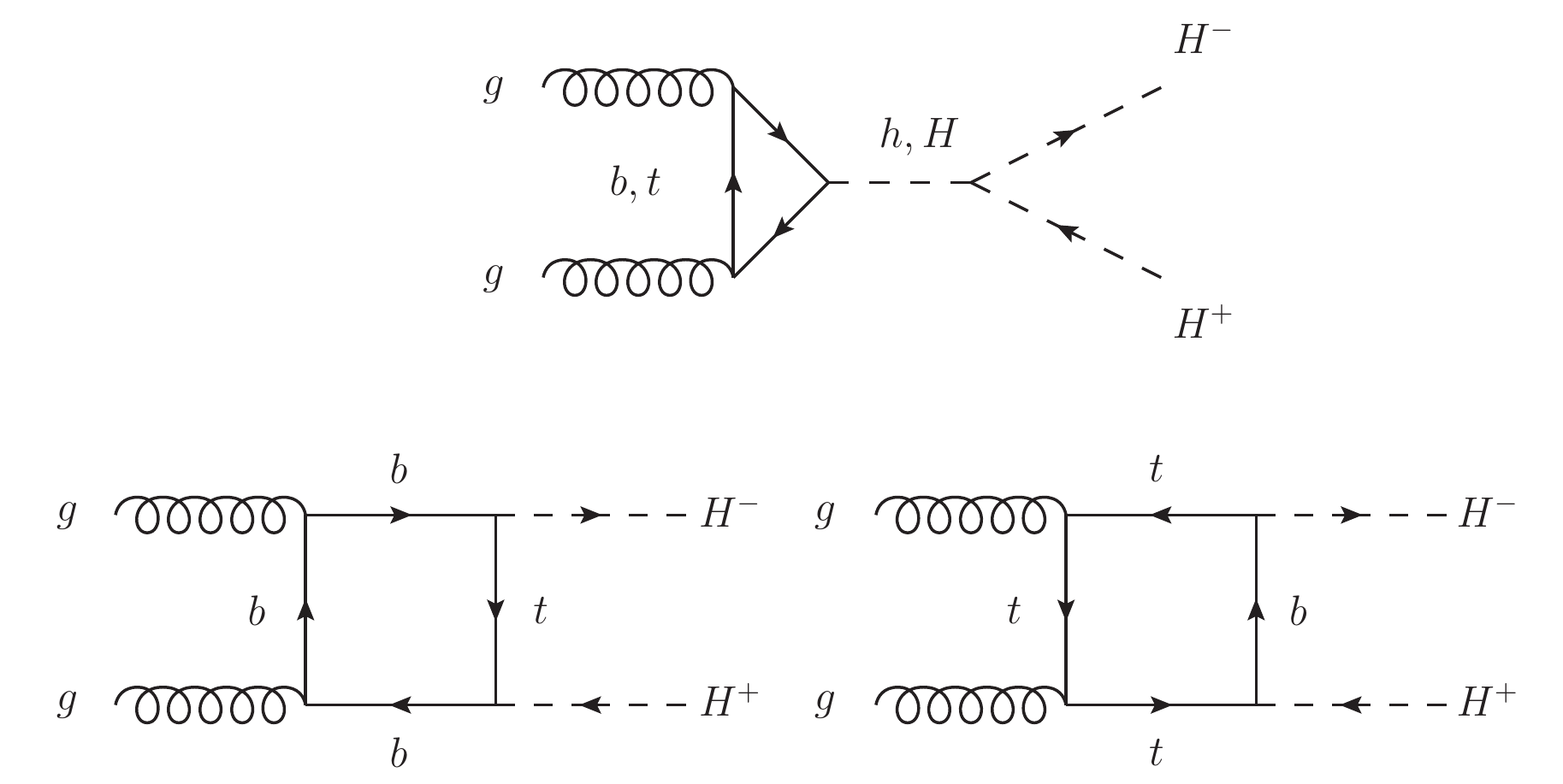}
\end{minipage}
\begin{minipage}[c]{5cm}
\includegraphics[scale=0.55]{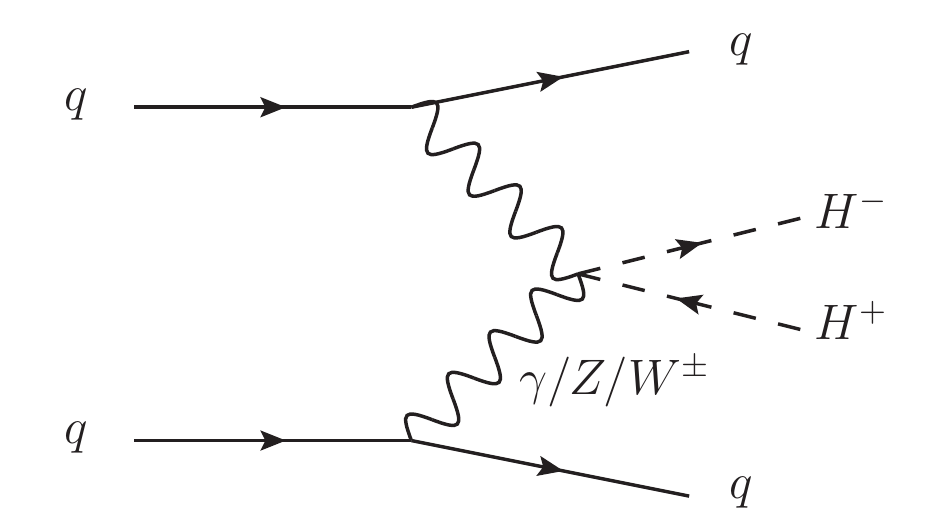}
\end{minipage}
\caption{Generic LO Feynman diagrams for the main production channels
  of charged Higgs boson pair production: Drell-Yan (upper left),
  bottom quark fusion (upper right), gluon fusion (lower left) and VBF
  (lower right).}
\label{fig:hh-feyndiag}
\end{figure}

The constraints on the charged Higgs bosons come from precision
measurements as well as from direct searches at LEP, the Tevatron and
the LHC. The former are the most stringent limits on the charged Higgs
boson mass, especially in a type II 2HDM. The global
fit~\cite{Flacher:2008zq} including all flavour observables was
updated in 2015 especially for the $B\to X_s^{}\gamma$ observable and
gives a lower limit of $M_{H^\pm_{}}^{}\geq
493$~GeV~\cite{Enomoto:2015wbn}, when including higher-order
corrections in ${\rm BR}(B\to X_s^{}\gamma)$
prediction~\cite{Misiak:2015xwa}. A type I 2HDM escapes this flavour
limit and the limits are much weaker, mainly coming from direct
searches at LEP, with $M_{H^\pm_{}}^{}\gsim 72.5$~GeV, while the limit
for a type II 2HDM is $M_{H^\pm_{}}^{}\gsim
80$~GeV~\cite{Abbiendi:2013hk}. The Tevatron
searches~\cite{Gutierrez:2010zz} are now superseeded by the LHC
results. For example LHC Run I data excludes charged Higgs boson
masses of $M_{H^\pm_{}}^{} < 140$~GeV in the MSSM with an $m_h^{\rm
  mod-}$ scenario~\cite{Aad:2014kga,Aad:2015typ}. The first Run II
results at 13 TeV have improved the direct limits in the high-mass
range, see Refs.~\cite{ATLAS:2016grc,ATLAS:2016qiq,CMS:2016qoa}.

\begin{figure}
\begin{center}
\includegraphics[scale=0.6]{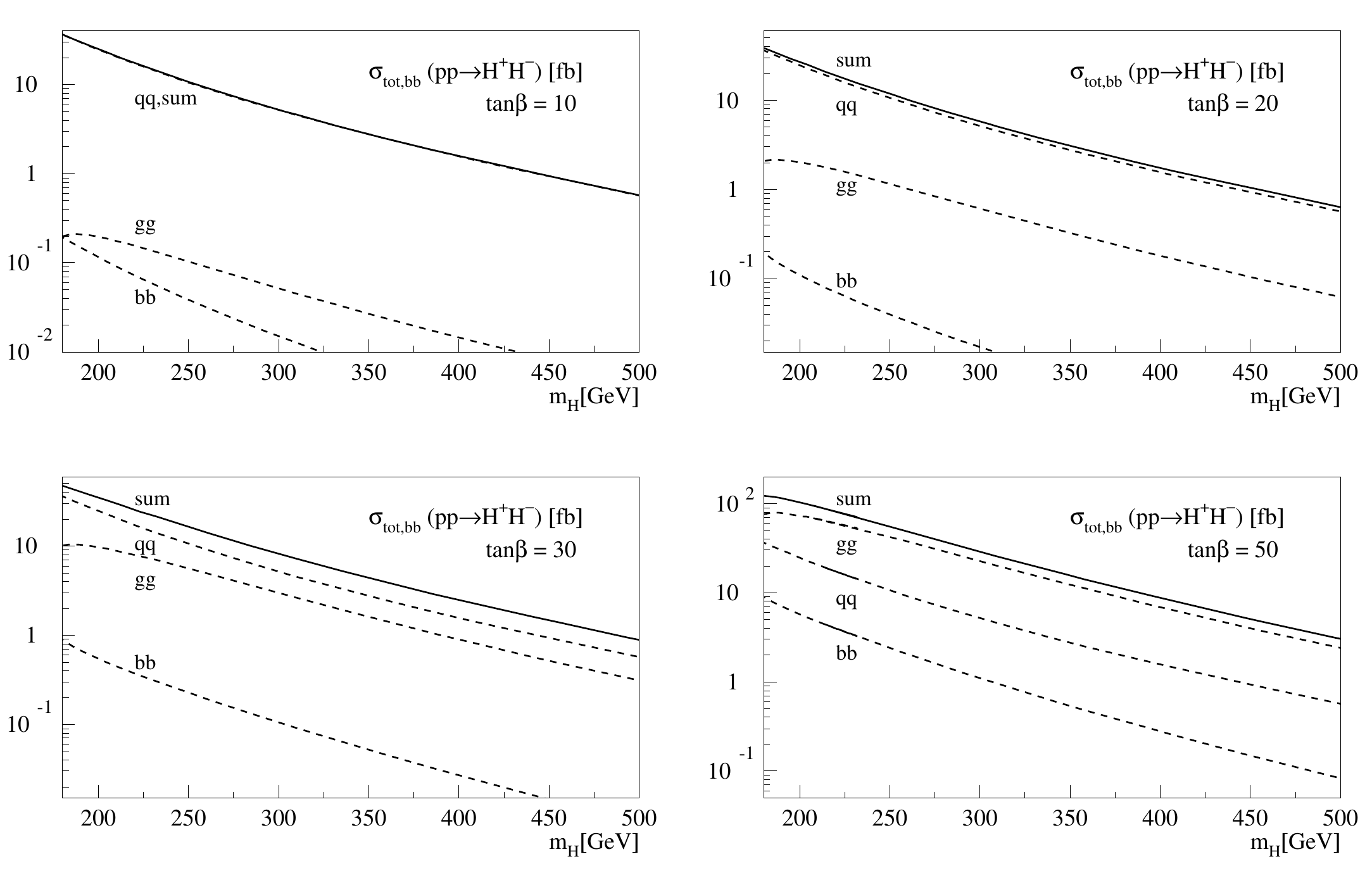}
\end{center}
\vspace{-5mm}
\caption{NLO QCD production cross sections (in fb) of $p p\to H^+_{}
  H^-_{}$ in the Drell-Yan channel (labelled as $qq$), the
  bottom-quark fusion channel (labelled as $bb$) and in the gluon
  fusion channel (labelled as $gg$) as a function of the charged Higgs
  boson mass $m_H^{}$ (in GeV). Four different values of $\tan\beta$
  are investigated. Taken from Ref.~\cite{Alves:2005kr}.}
\label{fig:hh-summary}
\end{figure}

There are three main classes of charged Higgs boson pair production at
the LHC. The dominant channel at low $\tan\beta \lsim 30$ is
Drell-Yan production, including the bottom-quark initiated channel
$b\bar{b}\to H^+_{} H^-_{}$, then gluon fusion which is dominant at
high $\tan\beta \gsim 50$ and finally vector boson fusion
(VBF) which is usually the second production channel. Generic leading
order (LO) Feynman diagrams are depicted in
Fig.~\ref{fig:hh-feyndiag}. The cross section are generically small as
exemplified in Fig.~\ref{fig:hh-summary} for Drell-Yan production
(including bottom-quark fusion) and gluon fusion at next-to-leading
order (NLO) in QCD, at different values of $\tan\beta$. This means
that high luminosity is required to observe the possible production of
a pair of charged Higgs bosons, except in the case of resonant
production, that could lead to a sizeable enhancement, see below for a
discussion of this possibility especially in a type I 2HDM.

\subsection{Drell-Yan and $b\bar{b}$ production}


The Drell-Yan process $q\bar{q}\to H^+_{} H^-_{}$ provides the largest
production channel at low $\tan\beta \lsim 30$. The LO cross section
was calculated in the
1980s~\cite{Lane:1983xr,Eichten:1984eu,Deshpande:1983xu}, then
reproduced and studied later in the
2000s~\cite{BarrientosBendezu:1999gp,Aoki:2011wd}. The NLO QCD
corrections induce a $+27\% (+17\%)$ increase at 14 TeV for
$M_{H^\pm_{}}^{}=160 (500)$~GeV, with $\sigma = 58
(0.23)$~fb~\cite{Alves:2005kr,Djouadi:1999ht}. The cross section has
no dependence on $\tan\beta$. The scale uncertainty is of order $\pm
25\%$. The supersymmetric (SUSY) QCD corrections have been calculated in
Ref.~\cite{Djouadi:1999ht} and are negligible.

The study of the additional bottom-quark fusion subprocess started in
the 2000s. The $b\bar{b}$ process suffers from an
ambiguity: $b$--quarks in the proton come from a gluon splitting and
the process could be viewed either as a direct bottom-quark fusion
production~\cite{BarrientosBendezu:1999gp,Moretti:2001pp} or as
$H^+_{}H^-_{}$ production in association with $b$-jets, $gg\to
H^+_{}H^-_{} b\bar{b}$~\cite{Moretti:2001pp,Moretti:2003px}. It was
found at first that direct production was one order of magnitude
larger. This was related to the key issue of defining the right
factorisation scale $\mu_F^{}$ as $\log(\mu_F^{}/m_b^{})$ terms that
are resummed in the $b$--parton picture could overestimate the cross
section if $\mu_F^{}$ were too big.

The calculation of the NLO QCD
corrections~\cite{Alves:2005kr,HongSheng:2005uy} solved the issue and
confirmed that the correct choice is $\displaystyle \mu_F^{} = \frac12
M_{H^\pm_{}}$. The QCD corrections add $\sim +55\%$ to the total cross
section in the 2HDM and the scale uncertainty is of order $\pm
25\%$~\cite{Alves:2005kr}. The cross section is dependent on
$\tan\beta$. In the MSSM the SUSY-QCD corrections are dominated by
negative resummed $\Delta_b^{}$-terms in the bottom quark Yukawa
coupling and depend strongly on the MSSM
spectrum~\cite{Alves:2005kr,HongSheng:2005uy}.

\subsection{The vector boson fusion channel}

\begin{figure}
\centering
\begin{minipage}[c]{7cm}
\hspace{-5mm}\includegraphics[scale=0.35,angle=90]{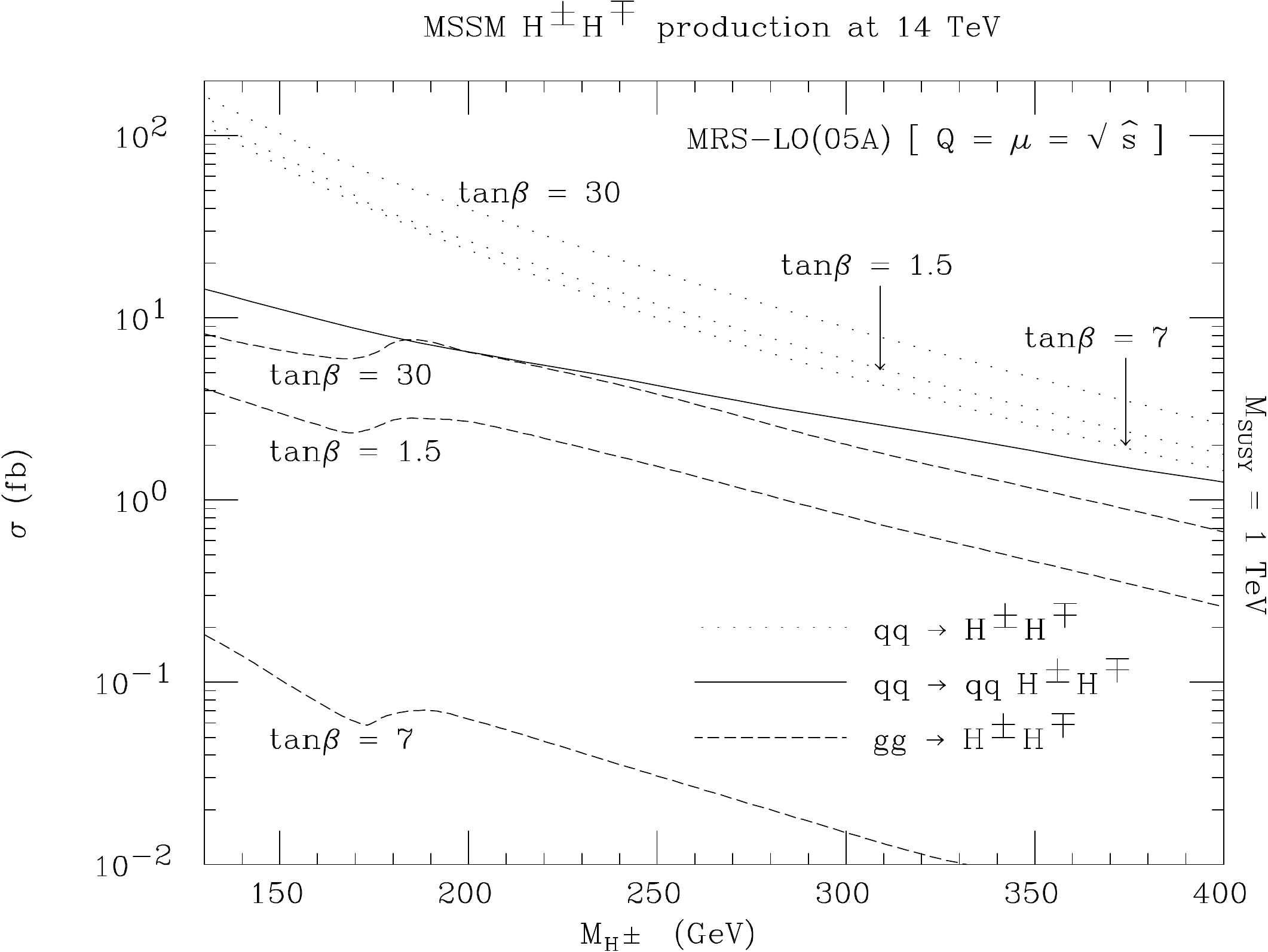}
\end{minipage}
\begin{minipage}[c]{7cm}
\hspace{5mm}\includegraphics[scale=0.35]{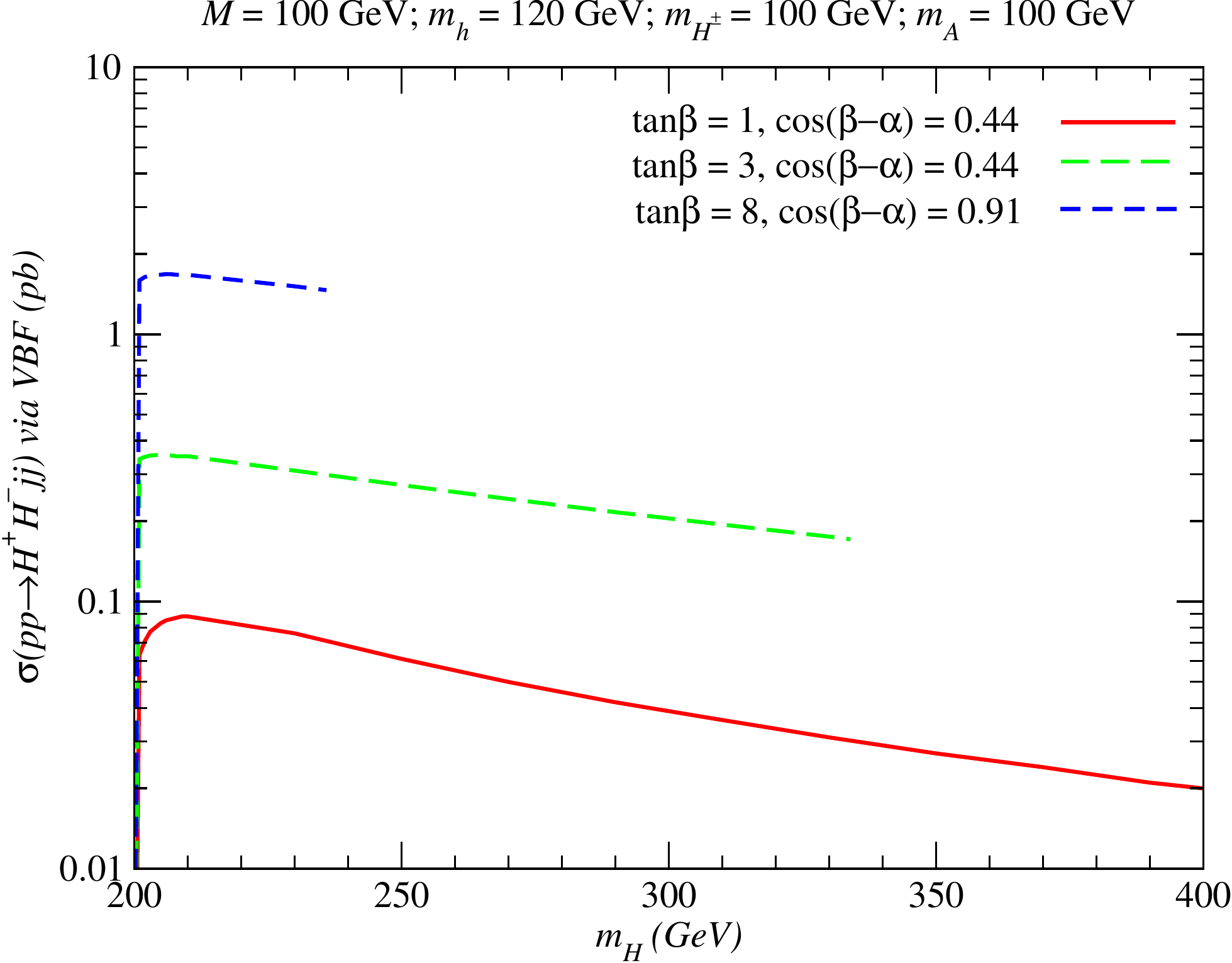}
\end{minipage}
\caption{Left: The Drell-Yan, VBF and gluon fusion $H^+_{} H^-_{}$
  production cross section at the 14 TeV LHC (in fb) as a function of
  the charged Higgs boson mass $M_{H^\pm_{}}^{}$ (in GeV) in the MSSM
  with a SUSY scale of 1 TeV, for different values of
  $\tan\beta$. Taken from Ref.~\cite{Moretti:2001pp}. Right: The VBF
  $H^+_{} H^-_{}$ production cross section (in pb) as a function of
  the heavy CP-even Higgs boson mass $m_{H}^{}$ (in GeV) in a 2HDM for
  different values of $\tan\beta$ and fixed values for $m_{h}^{},
  m_{H^\pm_{}}^{}$ and $m_{A}^{}$, complying with all known
  constraints on the model. Taken from Ref.~\cite{Aoki:2011wd}.}
\label{fig:hh-vbf}
\end{figure}

The VBF channel is usually the second most important production
channel at the LHC. The first approximate calculation was done in the
1980s for the 40 TeV Superconducting Super
Collider~\cite{Eboli:1987tu}. The LO full calculation in the MSSM for
the LHC was completed in 2001 for the total cross section as well as
for the differential distributions~\cite{Moretti:2001pp}, but has to
be taken with care as it uses an explicit quark mass in the MeV range
to regularize the collinear singularities due to photon exchange, that
is not consistent with the parton picture in QCD. The cross
section displays no dependence on $\tan\beta$ as exemplified in
Fig.~\ref{fig:hh-vbf} (left).

The LO cross section was recalculated for a general 2HDM (of any type)
and analysed in Ref.~\cite{Aoki:2011wd}. It was shown that resonant
effects in the $s$--channel due to heavy Higgs bosons are possible and
could enhance the cross section up to the picobarn level, see
Fig.~\ref{fig:hh-vbf} (right). The NLO QCD corrections are still
unknown.

\vspace{-3mm}\subsection{The gluon fusion channel}

We finish this review of the main production channels with the
one-loop initiated gluon fusion production that is known at LO. This
is the dominant channel for $\tan\beta \gsim 50$ and could feature
resonant effects. The earliest calculations can be traced back to the
1980s, with 2HDM approximate calculations including only the triangle quark
loop~\cite{Eichten:1984pd}. This was followed by the exact calculation
of the triangle quark loop, accompanied by the box loop in the heavy
quark limit in the 2HDM of type I and II~\cite{Foot:1987si}. The
calculation in the MSSM with loops of heavy squarks was completed at
the sime time~\cite{Willenbrock:1986ry}.

The process was studied again at the end of the 1990s and after the
completion of the full exact quark
contributions~\cite{Jiang:1997cg,Krause:1997rc} (the first calculation
being wrong and corrected by the second calculation) the full
calculation in the MSSM and in 2HDMs was completed by different groups
at nearly the same
time~\cite{BarrientosBendezu:1999gp,Jiang:1997cr,Brein:1999sy} and
later including CP violating effects~\cite{Jiang:2001ju}. The
matching with parton shower has been available since
2015~\cite{Hirschi:2015iia}.

\begin{figure}
\begin{center}
\begin{minipage}[c]{7cm}
\includegraphics[scale=0.45]{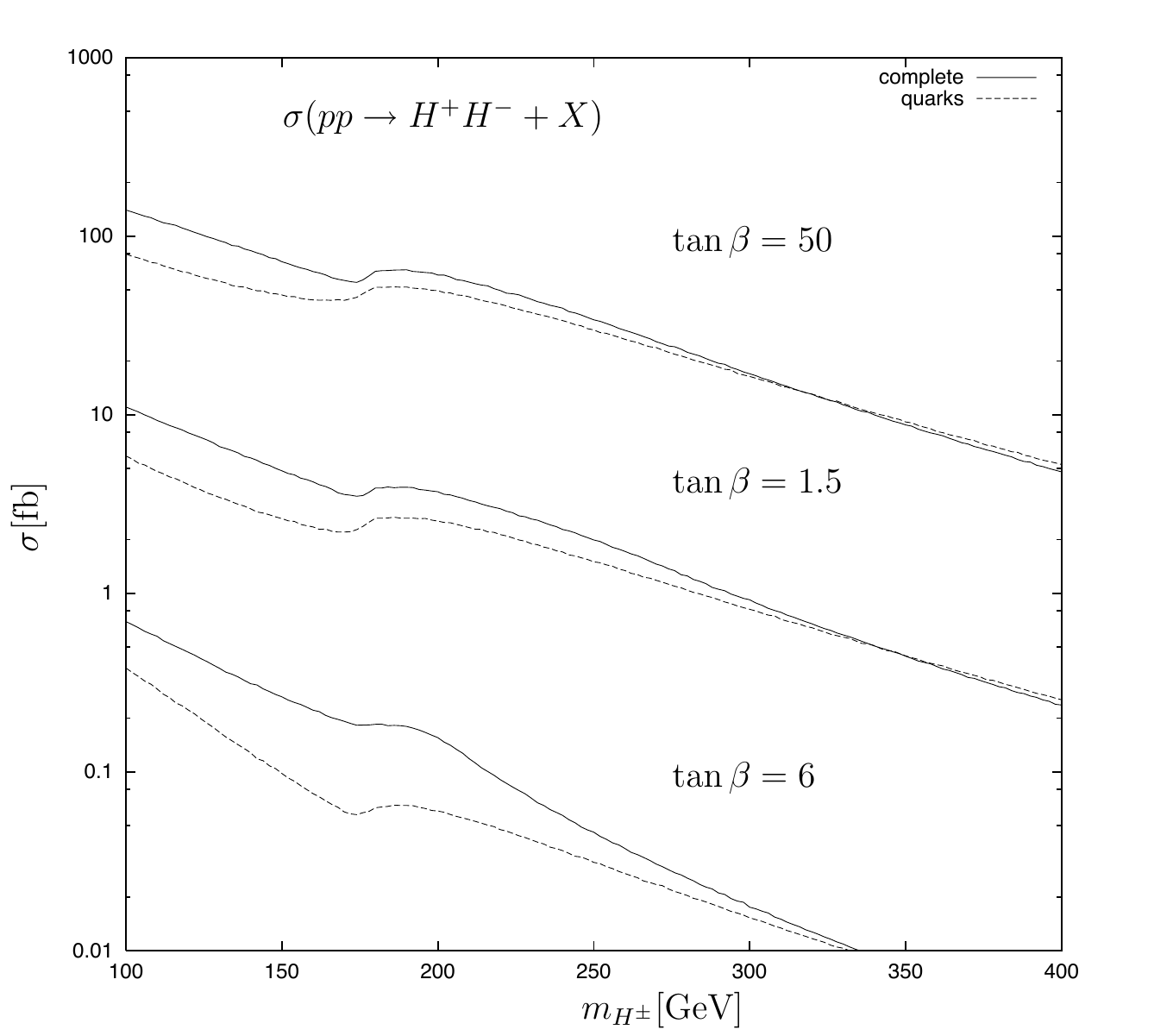}
\end{minipage}
\begin{minipage}[c]{7cm}
\includegraphics[scale=0.25]{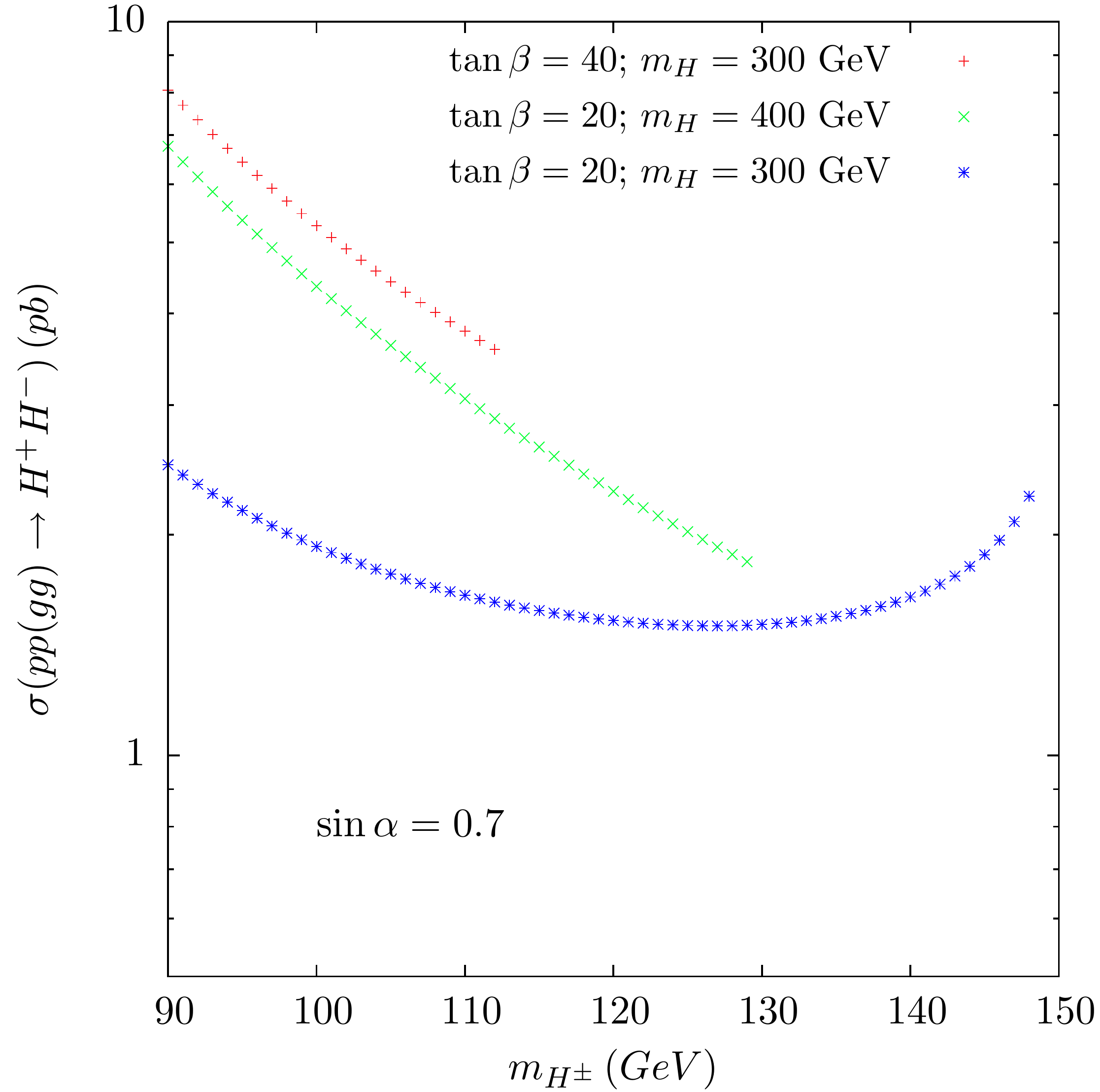}
\end{minipage}
\end{center}
\vspace{-4mm}
\caption{Left: The gluon fusion production cross section $p p\to
  H^+_{} H^-_{}$ at the 14 TeV LHC as a function of
  the charged Higgs boson mass $m_{H^\pm_{}}^{}$ (in GeV). Left: Cross
  sections (in fb) in the MSSM for different $\tan\beta$ values, with
  the quark contributions only (in dashed) and with the full
  contributions including squark loops (in solid). Taken from
  Ref.~\cite{Brein:1999sy}. Right: Cross sections (in pb) in a 2HDM
  with resonant effects for different values of $\tan\beta$ and
  of the heavy CP-even Higgs boson mass $m_{H}^{}$, complying with all
  known constraints on the model. Taken from Ref.~\cite{Aoki:2011wd}.}
\label{fig:hh-ggF}
\end{figure}

The squark loops lead to a large enhancement of the total cross
section~\cite{Brein:1999sy} as exemplified in Fig.~\ref{fig:hh-ggF}
(left). Depending on the MSSM parameters, the cross section can reach
200 fb at high $\tan\beta$. The scale uncertainty is estimated to be
of order $\pm 50\%$~\cite{Alves:2005kr}. In a 2HDM (in particular of type I)
resonant effects are possible and lead to pb cross sections as studied
in Ref.~\cite{Aoki:2011wd} and exemplified in Fig.~\ref{fig:hh-ggF}
(right). This makes this channel a very good probe of the
$h H^\pm_{} H^-_{}$ and $H H^\pm_{} H^-_{}$ couplings.

\section{Heavy neutrino impact on the triple Higgs coupling}

Since the Super-Kamiokande experiment in 1998 it is now experimentally
established that the neutrino flavours
oscillate~\cite{Fukuda:1998mi}. This calls for a BSM explanation as it
implies that neutrinos are not massless. One of the simplest
possibilities to explain neutrino masses is to add new fermionic heavy
gauge singlets that play the role of right-handed neutrinos in a
seesaw mechanism, see Ref.~\cite{Deppisch:2015qwa} for a review on
low-scale seesaw phenomenology at colliders.

We present a study of the impact of these new heavy neutrinos on the
triple Higgs coupling by considering a simplified 3+1 model where the
SM is minimally modified to account for 3 light massive Dirac
neutrinos and one heavy sterile Dirac neutrino. We recall that the
experimental prospects we take for the sensitivity to
$\lambda_{HHH}^{}$ are the following: $\sim 35\%$ at the
high-luminosity LHC when combining the $\sim 50\%$ sensitivity taken
from Ref.~\cite{CMS:2015nat}; $\sim 10\%$ at the ILC at 1 TeV with 5
ab$^{-1}_{}$~\cite{Fujii:2015jha}; and finally $\sim 5\%$ sensitivity
at the FCC-hh when combining the $\sim 8\%$ per experiment with 3
ab$^{-1}$ taken from Ref.~\cite{He:2015spf}.

\subsection{The simplified model and its constraints}

To illustrate the impact of a new, TeV scale fermion we use a
simplified model with 3 light active and one heavy sterile neutrinos
$n_i^{}$, parametrized by the masses $m_i^{}$ and the active-sterile
mixing matrix $B_{j k}^{}$, $i,k=1\dots 4$ and $j=1\dots 3$. The
Lagrangian contains the following neutrino interactions,
\begin{align}
\mathcal{L}_{n} =
  &  -\frac{g_2^{}}{\sqrt{2}} \bar{\ell}_i^{} \gamma^\mu_{} W^-_\mu
    B_{i j}^{} P_L^{} n_j^{} + {\rm h.c.}\nonumber\\
   & -\frac{g_2^{}}{2 \cos\theta_W^{}} \bar{n}_i^{} \gamma^\mu_{}
     Z_\mu^{} (B^\dagger_{} B)_{i j}^{} P_L^{} n_j^{}
     -\frac{g_2^{}}{2 M_W^{}} \bar{n}_i^{}  H (B^\dagger_{} B)_{i j}^{}
     \left(m_{n_i^{}}^{} P_L^{} + m_{n_j^{}}^{} P_R^{}\right) n_j^{}.
\end{align}
The active-sterile mixing matrix $B$ is build from the PMNS
matrix~\cite{Pontecorvo:1957cp,Maki:1962mu} extended to 4 neutrinos
with no CP violation in the neutrino sector. 

The new heavy neutrino generates new triangle one-loop contributions
for the triple Higgs coupling $\lambda_{HHH}^{}$ and lead to
modifications in the Higgs and weak bosons self-energies. Amongst the
experimental constraints that are applicable to the model, electroweak
precision observables (EWPO) are the
strongest~\cite{delAguila:2008pw,deBlas:2013gla}. Constraints coming
from neutrinoless $\beta$-decay do not apply and flavour
violating Higgs decays are less constraining than EWPO. We are lead
to the following constraints on the active-sterile mixing matrix,
\begin{align}
B_{e 4}^{} \leq 0.041,\ \ B_{\mu 4}^{} \leq 0.030,\ \ B_{\tau 4}^{} \leq 0.087.
\end{align}
We also apply theoretical constraints on the perturbativity of the
triple Higgs coupling as well as on the width of the heavy neutrino,
\begin{align}
\left(\frac{g_2^{} m_{n_4^{}}^{}}{2 M_W^{}} {\rm max}
  \left|(B^\dagger_{}B)_{i 4}^{}\right|\right)^3_{} < 16\pi (2\pi),\ \
  \Gamma_{n_4^{}}^{} \leq 0.06 m_{n_4^{}}^{}.
\end{align}
$2\pi$ stands for a tighter perturbativity bound roughly equivalent to
a two-loop analysis carried in the SM.

\subsection{Numerical results}

\begin{figure}
\begin{center}
\begin{minipage}[c]{7cm}
\includegraphics[scale=0.6]{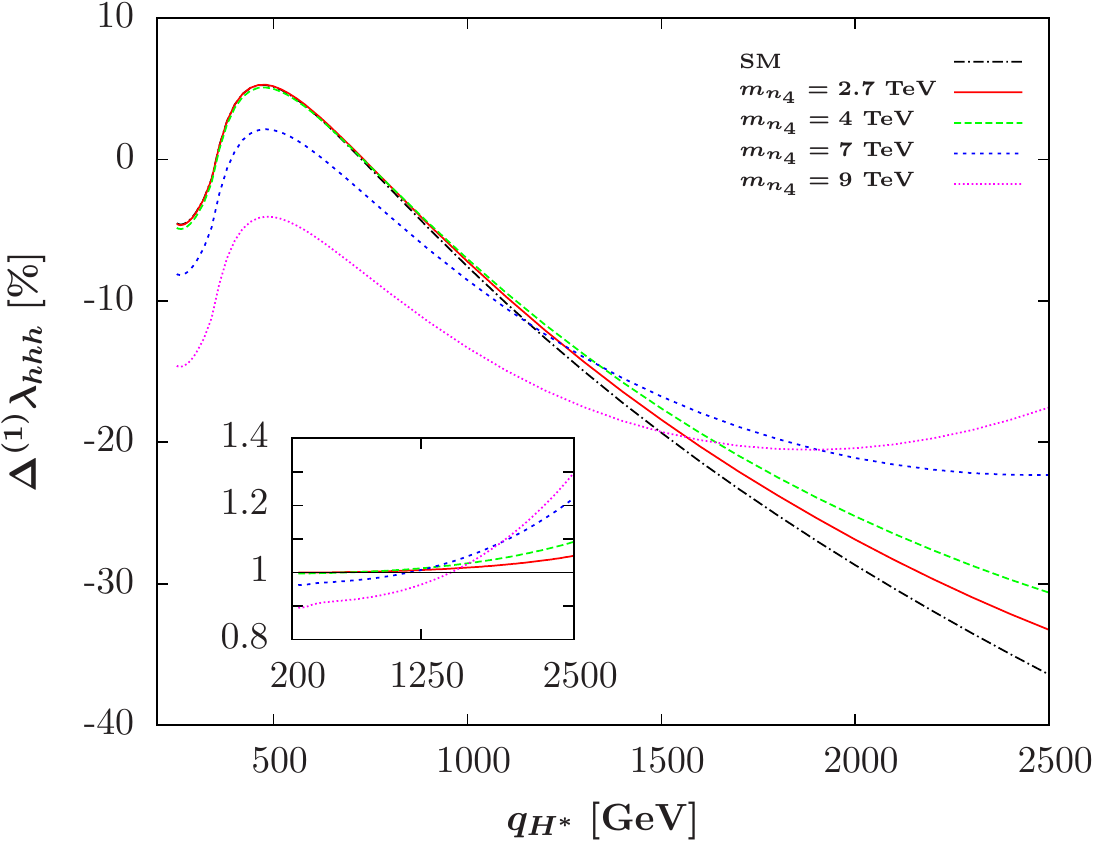}
\end{minipage}
\begin{minipage}[c]{7cm}
\includegraphics[scale=0.6]{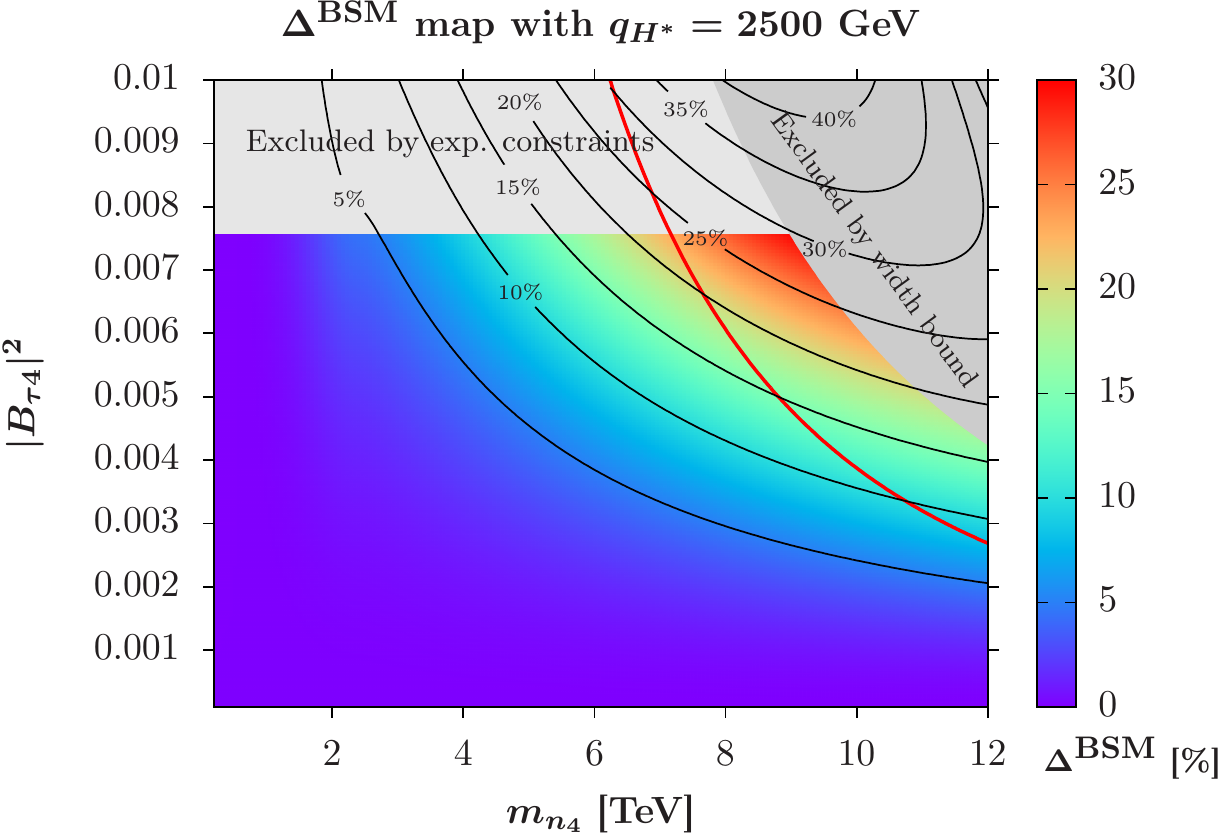}
\end{minipage}
\end{center}
\vspace{-4mm}
\caption{ Left: One-loop corrections to $\lambda_{HHH}^{}$ (in $\%$) as
  a function of the off-shell Higgs momentum $q_H^*$ (in GeV) of the
  splitting $H(q_H^*)\to H H$, with the neutrino parameter $B_{\tau
    4}^{} = 0.087$. Right: Contour map of the neutrino corrections
  $\Delta^{\rm BSM}_{}$ (in $\%$) as a function of the two neutrino
  parameters $m_{n_4^{}}^{}$ (in GeV) and $|B_{\tau 4}^{}|^2_{}$ at a
  fixed $q_H^*=2.5$ TeV. Both figures are taken from
  Ref.~\cite{Baglio:2016ijw}.}
\label{fig:neutrinos}
\end{figure}

To illustrate the impact of the heavy neutrino in our simplified model
we display in Fig.~\ref{fig:neutrinos} (left) the dependence of the
full one-loop corrections (including the SM contributions) on the
momentum $q_H^*$ of the off-shell Higgs splitting in two on-shell
Higgs bosons $H(q_H^*)\to H H$, for different heavy neutrino mass
values $m_{n_4}^{}$. We assume the maximal allowed value for $B_{\tau
  4}^{} = 0.087$ while $B_{e 4}^{} = B_{\mu 4}^{}=0$. Similar plots
are obtained for the other maximally allowed
mixings. $m_{n_4^{}}^{}=2.7$ TeV corresponds to the choice of a
neutrino Yukawa coupling equal to the top quark Yukawa coupling,
$m_{n_4^{}}^{}=7$ TeV is the maximal mass allowed by the tight
perturbativity bound while $m_{n_4^{}}^{}=9$ TeV is the maximal value
allowed by the width bound. The largest positive one-loop correction
is obtained at $q_H^* \simeq 500$ GeV and becomes smaller for a
heavier neutrino (eventually going negative). At large $q_H^* \simeq
2.5$ TeV the largest negative one-loop correction is obtained and the
heavier the neutrino is, the larger the correction becomes.

In Fig.~\ref{fig:neutrinos} (right) we display the contour map of the
genuine correction $\Delta^{\rm BSM}_{}$ (in percent) due to the heavy
neutrino, in the plane $(m_{n_4^{}}^{},|B_{\tau 4}^{}|^2_{})$. The
off-shell Higgs boson momentum is fixed to $q_H^*=2.5$ TeV. The red
line displays the tight perturbativity bound. We obtain corrections
as large as $\sim +30\%$ when taking into account the experimental and
theoretical constraints, at the limit of the high-luminosity
LHC and clearly visible at the ILC or at the FCC-hh. Again a similar
behaviour can be obtained for the active-sterile mixings $B_{e 4}^{}$
and $B_{\mu 4}^{}$.

\section{Outlook}

The production of a Higgs boson pair is one of the main goals of the
high-luminosity run of the LHC and of the future colliders, in order
to ultimately measure the triple Higgs coupling. Assessing the effects
of BSM models on the Higgs sector and in particular on the triple
Higgs coupling is thus an essential task.

As an illustrative example of such studies we have presented in a
first part a review of the current status for charged Higgs pair
production at the LHC in the 2HDM and the MSSM, focusing on the main
production channels: Drell-Yan production including bottom-quark
effects in the initial state, vector boson fusion and gluon
fusion. While the dominant channel at low $\tan\beta$, Drell-Yan
production, is known at NLO in (SUSY-) QCD, vector boson and
gluon fusions are only known at LO. Precision may be improved in
particular in the vector boson fusion channel.

We have presented in a second part the BSM effect induced by a heavy
neutrino on the triple Higgs coupling. Using a simplified model with
one heavy neutrino, effects as large as $+30\%$ can be found, that are
clearly measurable at future colliders. The triple Higgs coupling
is thus a new observable for neutrino physics that could be used to
probe neutrino mass regimes hard to access otherwise and would
constraint the active-sterile mixing.

\acknowledgments{
The author would like to warmly thank the organisers for the
invitation and for the very nice atmosphere of this
workshop. Discussions with Michael Spira are also acknowledged. He 
acknowledges the support from the Institutional Strategy of the
University of T\"ubingen (DFG, ZUK 63) and from the DFG Grant JA
1954/1.}


\bibliographystyle{JHEP-tweaked}
\bibliography{charged2016-bib}

\end{document}